\documentclass[%
 reprint,
superscriptaddress,
 amsmath,amssymb,
 aps,
prb,
floatfix,
]{revtex4-1}

\usepackage{graphicx}
\usepackage{dcolumn}
\usepackage{bm}
\usepackage{epstopdf}
\usepackage{natbib}
\usepackage{color}
\usepackage[hidelinks, colorlinks=true, urlcolor=blue, linkcolor=blue]{hyperref}
\renewcommand{\(}{\left(}
\renewcommand{\)}{\right)}
\begin{document}

\preprint{APS/123-QED}

\title{Tunable Graphene Metasurface Reflectarray for Cloaking, Illusion and Focusing}

\author{Sudipta Romen Biswas}
\affiliation{Department of Electrical \& Computer Engineering, University of Minnesota, Minneapolis, Minnesota 55455, USA}
\author{Cristian E Guti\'errez}
\affiliation{Department of Mathematics, Temple University, Philadelphia, Pennsylvania 19122, USA}
\author{Andrei Nemilentsau}
\affiliation{Department of Electrical \& Computer Engineering, University of Minnesota, Minneapolis, Minnesota 55455, USA}
\author{In-Ho Lee}
\affiliation{Department of Electrical \& Computer Engineering, University of Minnesota, Minneapolis, Minnesota 55455, USA}
\author{Sang-Hyun Oh}
\affiliation{Department of Electrical \& Computer Engineering, University of Minnesota, Minneapolis, Minnesota 55455, USA}
\author{Phaedon Avouris}
\affiliation{IBM T. J. Watson Research Center, Yorktown Heights, New York 10598, USA}
\author{Tony Low}
\email{tlow@umn.edu}
\affiliation{Department of Electrical \& Computer Engineering, University of Minnesota, Minneapolis, Minnesota 55455, USA}

\begin{abstract}
We present a graphene-based metasurface that can be actively tuned between different regimes of operation, such as anomalous beam steering and focusing, cloaking and illusion optics, by applying electrostatic gating without modifying the geometry of the metasurface. The metasurface is designed by placing graphene nano-ribbons (GNRs) on a dielectric cavity resonator, where interplay between geometric plasmon resonances in the ribbons and Fabry-Perot resonances in the cavity is used to achieve 2$\pi$ phase shift. As a proof of the concept, we demonstrate that wavefront of the field reflected from a triangular bump covered by the metasurface can be tuned by applying electric bias so as to resemble that of bare plane and of a spherical object. Moreover, reflective focusing and change of the reflection direction for the above-mentioned cases are also shown.
\end{abstract}

\maketitle
\section{Introduction}
Gradient metasurface is a planar arrangement of sub-wavelength scatterers of different shapes and sizes designed to structure wavefronts of reflected or transmitted optical beams by means of spatially varying optical response\cite{holloway2012,yu2014,jacob2017,kildishev2013planar,meinzer2014plasmonic}. Light interaction with the metasurfaces defies conventional laws of geometrical optics, such as the Snell's law or the law of reflection, and reveals a variety of non-trivial physical effects useful for practical applications. Particularly, efficient beam steering of the incident light in reflection and/or transmission modes was reported for metasurfaces operating both in narrow \cite{yu2011,aieta2012} and broad \cite{ni2012,sun2012,li2015,nemilentsau2017} frequency ranges. Moreover, pronounced polarization dependence of steering directions and/or amplitudes of beams deflected by metasurfaces was demonstrated \cite{farmahini2013,pfeiffer2013,pors2013,yin2013,wu2014,arbabi2015,shaltout15} thus paving the way for creating ultrathin optical polarizers, quarter and half wave plates\cite{yu2012broadband,zhao2013tailoring,ding2015broadband}. Great deal of attention has also been devoted to developing of viable alternatives to conventional focusing devices in transmission \cite{memarzadeh2011,aieta2012,chen2012, monticone2013, ni2013} (lenses) and reflection \cite{li2012, pors2013a, veysi2015, ma2016, zhang2016,fan2017} (parabolic reflectors) geometries. In fact, reflectarrays allow for the implementation of parabolic phase gradient along a planar surface thus avoiding technologically complicated process of creating parabolic surfaces for reflected light.

Recently it has been realized that metasurfaces can replace transformation optics\cite{li2008,lai2009,fleury2014} when it comes to implementing efficient cloaking devices. The essence of optical cloaking is to surround the object to be hidden by a material with carefully designed spatially varying dielectric permittivity (optical cloak) so that far-field radiation pattern of the object-cloak system is as close as possible to that of empty space. Efficient hiding of 2D and 3D bumps by metasurface carpet cloaks was reported \cite{estakhri2014, estakhri2015,orazbayev2015,yang2016,tao2016,cheng2016}. The advantage of the metasurface based cloaking is that control of the polarization, phase and amplitude of the wave reflected by a cloaked object can be achieved \cite{yang2016full} without modifying all the components of permittivity and permeability tensors which is required when using the transformation optics approach.

The operational characteristics (angle of beam steering, focal distance, angular efficiency, losses etc.) of optical devices based on metasurfaces designed using conventional dielectric or metal materials is typically predefined by the metasurface geometry and cannot be changed on-the-fly during the device operation. This might be a significant limitation when tuning of device characteristics is essential for the device operation, particularly, tunable steering angle for optical switches. Attempts to overcome this limitation using gate-tunable conducting oxides \cite{huang2016}, temperature-tunable nematic liquid crystals \cite{sautter2015}  or strain tunable elastic polymers \cite{kamali2016} as metasurface building blocks were reported. Graphene plasmonic resonators\cite{low2014graphene,grigorenko2012graphene,garcia2014graphene,bludov2013primer,malard2009raman,avouris20172d,christensen2011graphene} provides viable alternative\cite{fallahi2012,carrasco2013,carrasco2015,sherrott2017} to design of active metasurface that can be tuned by applying gate voltage. \textcolor{black}{Dynamic tuning of Fermi energy in graphene plasmonic structures has been reported for optical switching\cite{yu2015resonant} and infrared beam steering via acoustic modulation\cite{chen2014infrared}}. Active tuning of steering angle using graphene based metasurfaces operating in reflection regime was reported\cite{yatooshi2015,carrasco2015}.

The gradient metasurfaces are typically designed in order to perform a particular specialized task, such as tuning, focusing or cloaking. In this paper, we demonstrate that it is possible to design a versatile active metasurface using gate-tunable graphene ribbons\cite{yan2013damping,ju2011graphene} on an arbitrary substrate surface, which is capable of performing each of the above-mentioned specialized tasks depending on the electric bias profile across the surface of the metasurface, i.e. without changing the metasurface geometry. Particularly, we demonstrate that far-field distribution of the electric field of the wave reflected from a bump covered by such a metasurface can resemble either that of bare plane (cloaking case) or that of an object of a different shape (illusion), depending on the applied bias. In addition, we show that such wavefront engineering - as anomalous reflection and focusing - can also be achieved in conjunction with cloaking and illusion.

In what follows, we discuss general metasurface design strategy in Section II, followed by theoretical and simulation results for the above-mentioned functionalities in Section III-V. Lastly, we end with some general discussions on experimental realization and performance issues of the device in Section VI.

\section{Design of the Metasurface}
\begin{figure}[t]
\centering
\includegraphics[width=0.5\textwidth,scale=0.5]{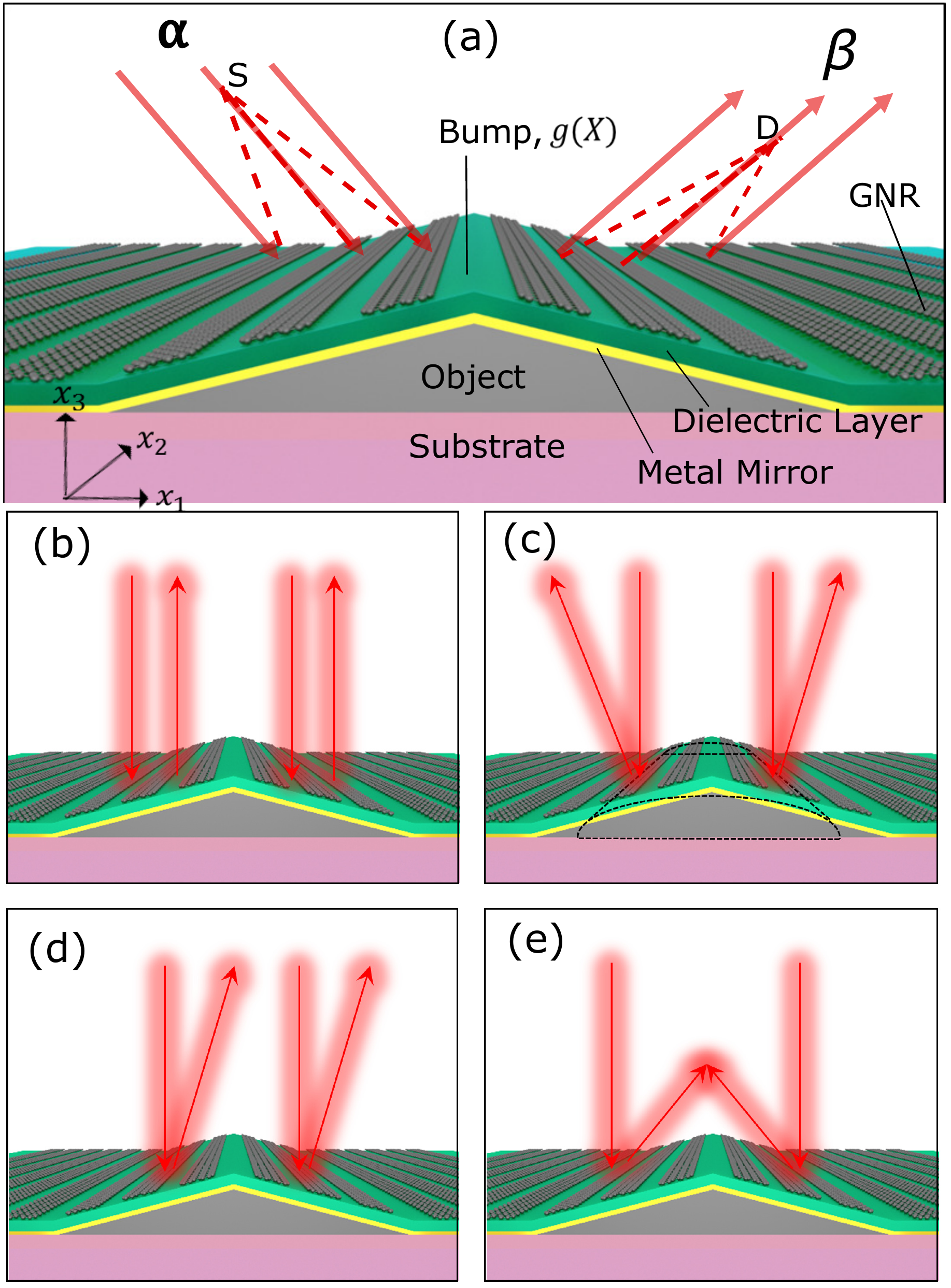}
\caption{Illustration of the metasurface design and applications. (a) shows structure of the GNR array metasurface covering a non-planar surface. $\bm{\alpha}$ and $\bm{\beta}$ are directions of incident and reflected rays. (b)-(e) show depiction of different reflection jobs discussed in the work. They are (b) cloaking with specular reflection, (c) illusion optics, (d) cloaking with anomalous reflection and (e) reflective focusing. }
\label{fig:fig1}
\end{figure}
Fig.\,\ref{fig:fig1}a shows a schematic of the graphene based metasurface device. In general, the metasurface can be implemented on a non-planarized surface. At the desired frequencies, mid-infrared light incident on the metasurface can be  reflected in non-trivial fashion to achieve various functionalities. For example, the light can be reflected as if the surface is planar (see Fig.\,\ref{fig:fig1}b) or disguised as a different surface morphology (see Fig.\,\ref{fig:fig1}c). The former is often referred to as cloaking in the literature \cite{ni2015ultrathin,chen2011mantle}, while the latter as illusion optics\cite{lai2009illusion}. The light can also be anomalously reflected to far field as plane wave in predetermined direction (see Fig.\,\ref{fig:fig1}d), or onto a focal point at the near field (see Fig.\,\ref{fig:fig1}e), all achieved on a non-planar substrate.

The general implementation of these various reflection modes can be achieved with the appropriate phase discontinuities, $\phi$,  at the graphene metasurface. The phase discontinuity for any arbitrary reflection beam wavefront can be derived from ray optics arguments. \textcolor{black}{Let us consider a general surface in 3D space, with coordinates of a point $P$ on the surface defined as
\begin{equation} \label{Eq:point}
P = (u_1, u_2, u_3),
\end{equation}
where $u_1 = x_1$, $u_2 = x_2$, $u_3 = g(X)$, $X = (x_1, x_2)$ (see Fig. \ref{fig:fig1}a). The normal to the metasurface, $\nu(P)$, is
\begin{equation} \label{eq:normal}
\bm{\nu}(P)=\frac{\(-\nabla g(X),1\)}{\sqrt{1+|\nabla g(X)|^2}}.
\end{equation}
Suppose, $\bm{\alpha}(P)$ and $\bm{\beta}(P)$ are unit direction vectors for incident and reflected waves}. In absence of any phase discontinuity along the surface, we can write the vector form of conventional Snell's law as\cite{luneburg1964mathematical}
\begin{equation}\label{eq:law of reflection}
\bm{\alpha}(P)\times\bm{\nu}(P)=\bm{\beta}(P)\times\bm{\nu}(P)
\end{equation}
which is equivalent to $(\bm{\alpha}(P)-\bm{\beta}(P))\times\bm{\nu}(P)=0$ i.e $\bm{\alpha}(P)-\bm{\beta}(P)$ is parallel to $\bm{\nu}(P)$. Therefore, we can write\cite{gutierrez2017general,luneburg1964mathematical}
\begin{equation*}
\bm{\alpha}(P)-\bm{\beta}(P)=\lambda\,\bm{\nu}(P)
\end{equation*}
where $\lambda$ is a scalar factor, $\lambda\in \mathbb R$.
When we have a phase discontinuity, given by a function $\phi$,  defined in the neighborhood of the surface, the generalized law of reflection in vector form\cite{gutierrez2017general} is given by (Appendix: \ref{app1})
\textcolor{black}{
\begin{equation}\label{eq:generalized law of reflection}
\bm{\alpha}(P)-\bm{\beta}(P)=\frac{\nabla \phi(P)}{{{k_0}}}+\lambda\,\bm{\nu}(P),
\end{equation}}
where $k_0$ is the free space wave number. Based on the desired operation, one would stipulate the required scattering beams $\bm{\alpha}$ and $\bm{\beta}$, and starting from Eq.\,\eqref{eq:generalized law of reflection} we can calculate the respective phase profiles $\phi$. We defer these calculations to sections III-V.

In practice, design of a phase control metasurface involves two steps\cite{scheuer2017metasurfaces}. First, a phase profile or phase mask for the desired wavefront modification is calculated and then individual pixels of the phase profile, which locally tailor the phase of the impinging wave, are designed. The scattering phase is achieved with graphene plasmonic nanoribbons\cite{low2014graphene,grigorenko2012graphene,garcia2014graphene,bludov2013primer,avouris20172d}, whose plasmon resonance is tunable with doping or width. In this work, we fix the ribbon widths and vary the doping to achieve the desired phase $\phi$.

Fig.\,\ref{fig:fig1}a provides an illustration of the graphene nanoribbon based metasurface on a dielectric layer. There is a metal mirror below the dielectric layer separated at quarter wavelength distance from the graphene arrays. This maximizes the field at the graphene surface, hence enhancing light-matter interactions\cite{carrasco2015}. To have total control over wavefront, the phase shift along the metasurface needs to encompass the full $2\pi$ range. Graphene nanoribbon, with its Lorentzian-like response, provides a phase shift of only $\pi$. The interference between the graphene resonator and the Fabry-Perot cavity provides the extra phase shift to make the total range of phase variation very close to $2\pi$\cite{carrasco2015}. From the phase profile function, $\phi(P)$, which we derive in Sections III-VI, we will be able to assign the required phase to each respective nanoribbon.

In this work, graphene conductivity is described with the finite temperature Drude formula which accounts for the intraband optical processes,
\begin{equation}\label{eq:Drude}
\sigma \left( {{E_F}} \right) = \frac{{2{e^2}}}{{\pi {\hbar ^2}}}{k_B}T.\log \left[ {2\cosh \left( {\frac{{{E_F}}}{{2{k_B}T}}} \right)} \right]\frac{i}{{\omega  + i{\tau ^{ - 1}}}}.
\end{equation}
$E_F$ is the Fermi level of the nanoribbon, which is chosen according to the desired scattering phase, $\omega$ is the angular frequency taken to be equal to a free space wavelength of 22$\mu$m, $\tau$ is the graphene relaxation time, $e$ is the electronic charge, $T=300$K is the temperature. \textcolor{black}{While choosing the value of relaxation time, the fact that plasmon damping increases due to interaction with optical phonons from graphene and substrate should be considered\cite{yan2013damping}. In this work, we assume a free space wavelength of 22$\mu$m, which is significantly lower than the optical phonon energy ($\sim$0.2eV) in graphene. Moreover, we assume a substrate that does not have surface optical phonons at the operating frequency, so the choice of relaxation time $>$0.1ps to ensure availability of 2$\pi$ shift (see Appendix \ref{app2}) is justified. For example, CaF$_2$ is transparent in mid-infrared. We use a value of $\tau$ = 0.6ps\cite{yan2013damping}. For the dielectric layer, we assume a lossless refractive index of $n$=1.4 with thickness of 3.93$\mu$m corresponding to the quarter wavelength condition.}

Simulations are performed using Maxwell equation solver COMSOL Multiphysics\cite{comsol} RF Module. We model each graphene ribbon in terms of its 2D current density. For this, we need to translate the spatial phase profile into corresponding conductivity profile. First, we define the position of each ribbon by the coordinates of their centers. Then using the phase profiles $\phi$ derived in Sections III-V, we get the discrete phase values for the ribbons. Using these phase values, we can determine the corresponding Fermi energy ($E_F$) for individual ribbons.  Then, we get the required conductivity by putting the $E_F$ values in the Drude equation (Eq.\,\eqref{eq:Drude}). Finally in COMSOL, we put this spatial conductivity profile defined for each nanoribbon as the conductivity of the surface current densities. A fixed ribbon width of 500nm and inter-ribbon distance of 750nm are used. $E_F$ is varied between 0.15-0.8eV. Perfectly Matched Layer (PML) conditions are used at the simulation domain boundaries and the metal reflector is modeled with a Perfect Electric Conductor (PEC).

\section{Cloaking: Specular and Anomalous Reflection} \label{Sec3}
\textcolor{black}{In this section we derive the phase function, $\phi(P)$, required for cloaking with specular or anomalous reflected beams. We assume that metasurface is parametrized by \eqref{Eq:point},\eqref{eq:normal}. Following \eqref{eq:generalized law of reflection},  we seek $\phi$ such that the metasurface reflects all incident rays with direction $\bm{\alpha}$ into rays with direction $\bm{\beta}$, where $\bm{\alpha}$ and $\bm{\beta}$ are constant with respect to $P$. Taking double cross product of Eq.\,\eqref{eq:generalized law of reflection} with $\bm{\nu}(P)$ yields
\textcolor{black}{
\begin{align}\label{eq:double cross product}
0&=\bm{\nu}\times \left(\left(\bm{\alpha}-\bm{\beta}-\nabla\phi/k_0\right)\times \bm{\nu}\right)\nonumber\\
&=
\left(\bm{\alpha}-\bm{\beta}-\nabla\phi/k_0\right) - \left(\bm{\nu}\cdot \left(\bm{\alpha}-\bm{\beta}-\nabla\phi/k_0\right)\right)\,\bm{\nu}.
\end{align}}
We seek $\phi$ such that $\nabla \phi(P) = (\phi_{u_1}(P), \phi_{u_2}(P), \phi_{u_3}(P))$ is tangential to the surface, i.e. $\bm{\nu}\cdot \nabla \phi=0$. Here, and in rest of the paper, the notation $\phi_{u_i}(P)$ means derivative of $\phi(P)$ with respect to $u_i$.
Therefore from \eqref{eq:normal},\eqref{eq:double cross product} we obtain
\begin{equation}\label{eq:equation for nabla psi}
\nabla\phi(P)=k_0 \left(\bm{\alpha}-\bm{\beta} - \delta\,(-\nabla g(X),1) \right),
\end{equation}
where
\begin{equation} \label{eq:vector_cloaking}
\delta = \(\dfrac{(\bm{\alpha}-\bm{\beta})\cdot (-\nabla g(X),1)}{1+|\nabla g(X)|^2} \).
\end{equation}}

\textcolor{black}{Eq. \eqref{eq:equation for nabla psi} is a system of three differential equations for unknown phase function, $\phi(P)$, written in vector form (see Appendix \ref{app:coord} for coordinate form),  which can be reduced to two equations by taking into account that $\phi(P)$ is in fact a function of two variables, $x_1$, $x_2$ (see \eqref{Eq:point}). Using chain rule, we obtain
\begin{align}
\dfrac{\partial \phi }{\partial x_i}
&=
\dfrac{\partial \phi}{\partial u_i}+\dfrac{\partial \phi}{\partial u_3}\frac{\partial u_3}{\partial x_i}\notag\\
&=k_0\(\alpha_i-\beta_i +\(\alpha_3-\beta_3\)\,g_{x_i}(X)\), \label{eq:chain_rule}
\end{align}
where $i = 1,2$, $g_{x_i}(X) = \partial g(X)/\partial x_i$, and $\partial\phi/\partial u_i$ are defined by \eqref{eq:equation for nabla psi}. }
Integrating, we obtain the phase:
\begin{align}\label{eq:phase discontinuity constant reflection}
\phi\(X,g(X)\)&=k_0\(\(\alpha_1-\beta_1\)\,x_1+\(\alpha_2-\beta_2\)\,x_2\right.\nonumber \\
&+\left.\(\alpha_3-\beta_3 \)\,g(X)\)+C,
\end{align}
with $C$ an arbitrary constant. For 2D geometry, i.e where the equations are independent of $x_2$, the last equation can be written as
\begin{equation}\label{eq:phase discontinuity constant reflection_2D}
\phi(x_1)=k_0\(\(\alpha_1-\beta_1\)\,x_1+\(\alpha_3-\beta_3 \)\,g(x_1)\)+C.
\end{equation}

In terms of the incident angle $\theta_i$ and the reflection angle $\theta_r$, we have $\alpha=(-\sin\theta_i,-\cos\theta_i)$, $\beta=(-\sin\theta_r,\cos\theta_r)$. So in terms of $\theta_i$ and $\theta_r$, \eqref{eq:phase discontinuity constant reflection_2D} becomes
\begin{equation}\label{eq:phase discontinuity constant reflection_2D_angles}
\boxed{
\begin{aligned}
\phi(x_1)&=k_0\(\(\sin\theta_r-\sin\theta_i\)\,x_1\right.\\
&-\left.\(\cos\theta_r+\cos\theta_i\)\,g(x_1)\)+C.
\end{aligned}
}
\end{equation}

This is the general phase equation for cloaking. When $\theta_r=\theta_i$, this gives the phase for cloaking with specular reflection.
\begin{figure}[t]
\centering
\includegraphics[width=0.5\textwidth]{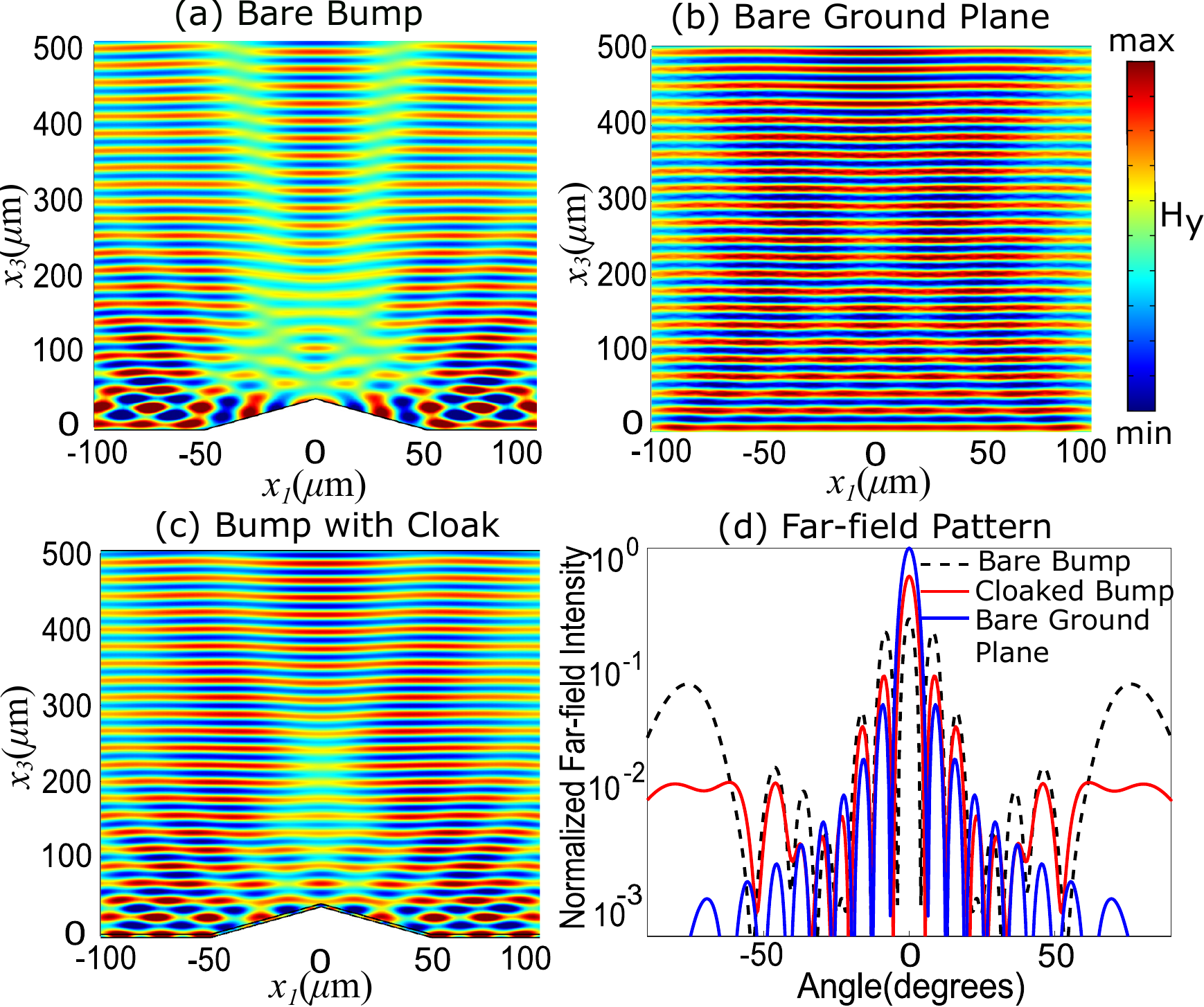}
\caption{Simulation results for cloaking with specular reflection. The cloaked object is a triangle shaped bump. (a),(b) and (c) show the scattered field plots for bare bump, ground plane and cloaked bump, while (d) shows corresponding far-field plots.}
\label{fig:fig2}
\end{figure}

Fig.\,\ref{fig:fig2} shows simulation results for the specular cloaking case. We have a triangular shaped bump with base length 100$\mu$m and height 40$\mu$m as the object to be cloaked. Results are shown for normal incidence of light. Fig.\,\ref{fig:fig2}a and \ref{fig:fig2}b show scattered field (magnetic field $H_y$) plots for the bare bump and bare ground plane respectively. Next, the bump is cloaked by the metasurface designed with the above mentioned $\phi$ and the scattered field plot is shown in  Fig.\,\ref{fig:fig2}c. Accompanying angle resolved far-field intensity plots are shown in log-scale in Fig.\,\ref{fig:fig2}d. As we can see, within the angular window of $\pm 40^{\circ}$, the angular-resolved intensity spectrum for the cloaked bump and bare ground plane far-field match very well. The presence of side lobes in the far-field for the bare ground plane can be attributed to the finiteness of simulation domain. If we increase the size of the simulation domain, both the main lobes and side lobes become narrower and ideally, with infinitely large simulation domain, we can expect only one narrow main lobe.
\begin{figure}[t]
\centering
\includegraphics[width=0.5\textwidth]{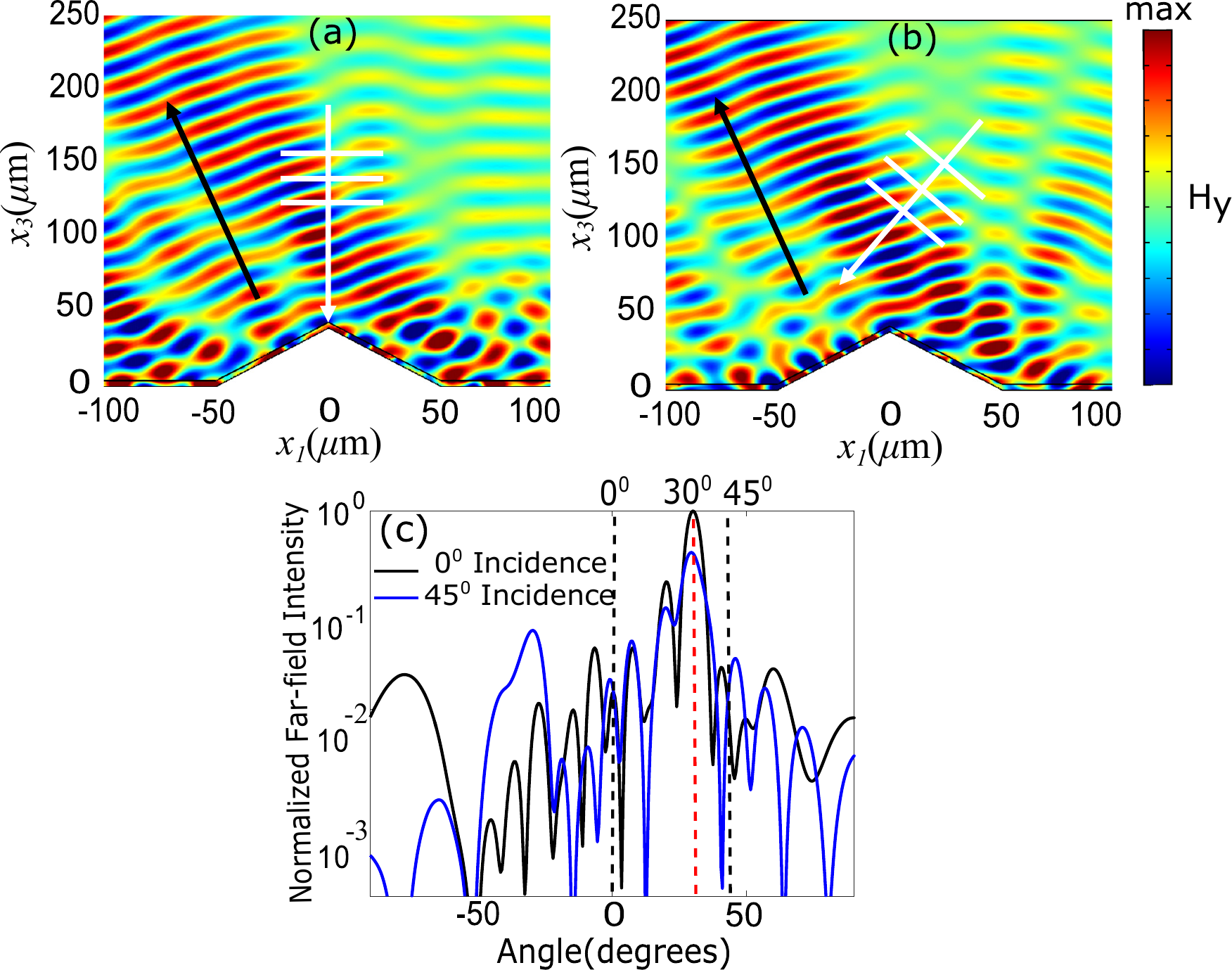}
\caption{Simulation results for cloaking with anomalous reflection. (a) and (b) show scattered field plots for normal and $45^{\circ}$ incidence respectively. For both cases the angle of reflection is designed to be $30^{\circ}$. Corresponding far-field plots are shown in (c).}
\label{fig:fig4}
\end{figure}

In similar fashion, we can also implement an extended version of the cloak, but with non-specular reflection angle.
Fig.\,\ref{fig:fig4}a demonstrates such implementation, designed with a $30^{\circ}$ angle of reflection off normal. In Fig.\,\ref{fig:fig4}a and \ref{fig:fig4}b, the scattered fields are shown for normal and $45^{\circ}$ angle of incidence, respectively. The white and black arrows show the incident and reflected wave directions. There are some distortions in the wavefronts predominantly due to specular reflections from the ground plane. In addition, we can also notice specular reflection on the right side of the bump. As we can see, the main beam is scattered at $30^{\circ}$ off normal per the design while power flow in specular directions ($0^{\circ}$ and $45^{\circ}$) are more than an order of magnitude smaller.

\section{Illusion Optics}
\textcolor{black}{Suppose that a surface $\Gamma'$ in 3D space is parameterized by a function $(X,f(X))$ and no phase discontinuity is given on $\Gamma'$. The reflection of the rays by such a surface is governed by the standard Snell's law of reflection,}
\begin{equation}\label{eq:standard snell for first surface}
\bm{\beta}(P')=\bm{\alpha} -2\,\(\bm{\alpha}\cdot \bm{\eta}(P')\)\,\bm{\eta}(P'),
\end{equation}
\textcolor{black}{where $P' = (X,f(X))$ is a point on $\Gamma'$, $\bm{\alpha}$, $\bm{\beta}(P')$ are the unit direction vectors for incident and reflected waves,  and $\bm{\eta}(P')=\dfrac{\(-\nabla f(X),1\)}{\sqrt{1+|\nabla f(X)|^2}}$ is the unit normal.}

\textcolor{black}{We consider another metasurface, $\Gamma$, parameterized by \eqref{Eq:point}, \eqref{eq:normal} and derive a phase discontinuity, $\phi(P)$, such that the metasurface $(\Gamma,\phi)$ does the same reflection job as the surface $\Gamma'$. That is, at each point $P$ the incident ray with unit direction $\bm{\alpha}$ is reflected into the ray with unit direction $\bm{\beta}(P')$ given in \eqref{eq:standard snell for first surface} }.
From \eqref{eq:generalized law of reflection} we then seek $\phi$ such that
\textcolor{black}{
\begin{equation}
\bm{\alpha}-\bm{\beta}(P') -  \frac{\nabla \phi(P)}{k_0}=\lambda \,\bm{\nu}(P)
\end{equation}
}

\textcolor{black}{As in Section \ref{Sec3}, making double cross product of this equation with $\bm{\nu}$ and assuming that $\nabla \phi\cdot \bm{\nu}=0$,} yields
\textcolor{black}{
\begin{align}
&\frac{\nabla \phi(P)}{k_0}  =\bm{\alpha}-\bm{\beta}(P')-\(\(\bm{\alpha}-\bm{\beta}(P')\)\cdot \bm{\nu}(P)\)\,\bm{\nu}(P) \notag \\
& =2\,\(\bm{\alpha}\cdot \bm{\eta}(P')\) \left\{\bm{\eta}(P')-\(\bm{\eta}(P')\cdot \bm{\nu}(P)\)\,\bm{\nu}(P)\right\}, \label{eq:gradient}
\end{align}
where we used \eqref{eq:standard snell for first surface} to obtain the second line. Equation  \eqref{eq:gradient} is a vector form of a system of three differential equations  (see Appendix \ref{app:coord} for coordinate form), which once again can be simplified using the chain rule
\begin{align}\label{eq:partial xi psi general}
\dfrac{\partial \phi}{\partial x_i}
&= \dfrac{\partial \phi}{\partial u_i}+\dfrac{\partial \phi}{\partial u_3}\frac{\partial u_3}{\partial x_i}\notag\\
&=
 2k_0\(\bm{\alpha}\cdot \bm{\eta}(P')\)\, \dfrac{g_{x_1}(X)-f_{x_1}(X)}{\sqrt{1+|\nabla f(X)|^2}} := A_i(X),
\end{align}
where $i = 1,2$. Integrating system of two differential equations, \eqref{eq:partial xi psi general}, we obtain (see Appendix \ref{app:integrate} for details):}
\begin{equation}\label{eq:formula for phase general case}
\phi\(X,g(X)\)=\int_a^{x_1}A_1(s,x_2)\,ds+\int_b^{x_2}A_2(a,t)\,dt +C.
\end{equation}
For the case where the configuration is independent of $x_2$, formula \eqref{eq:formula for phase general case} can be simplified to
\begin{equation}\label{eq:phase when configuration is independent of x2}
\boxed{
\phi(x_1)
=
2k_0\,\int_a^{x_1} \dfrac{\(-\alpha_1\,f'(s)+\alpha_3\)\(g'(s)-f'(s)\)}{1+f'(s)^2}\,ds+C.}
\end{equation}
Which gives the phase required to be applied along a surface $g(x_1)$ to mimic the reflection pattern of another surface $f(x_1)$.

\begin{figure}[t]
	\centering
	\includegraphics[width=0.49\textwidth]{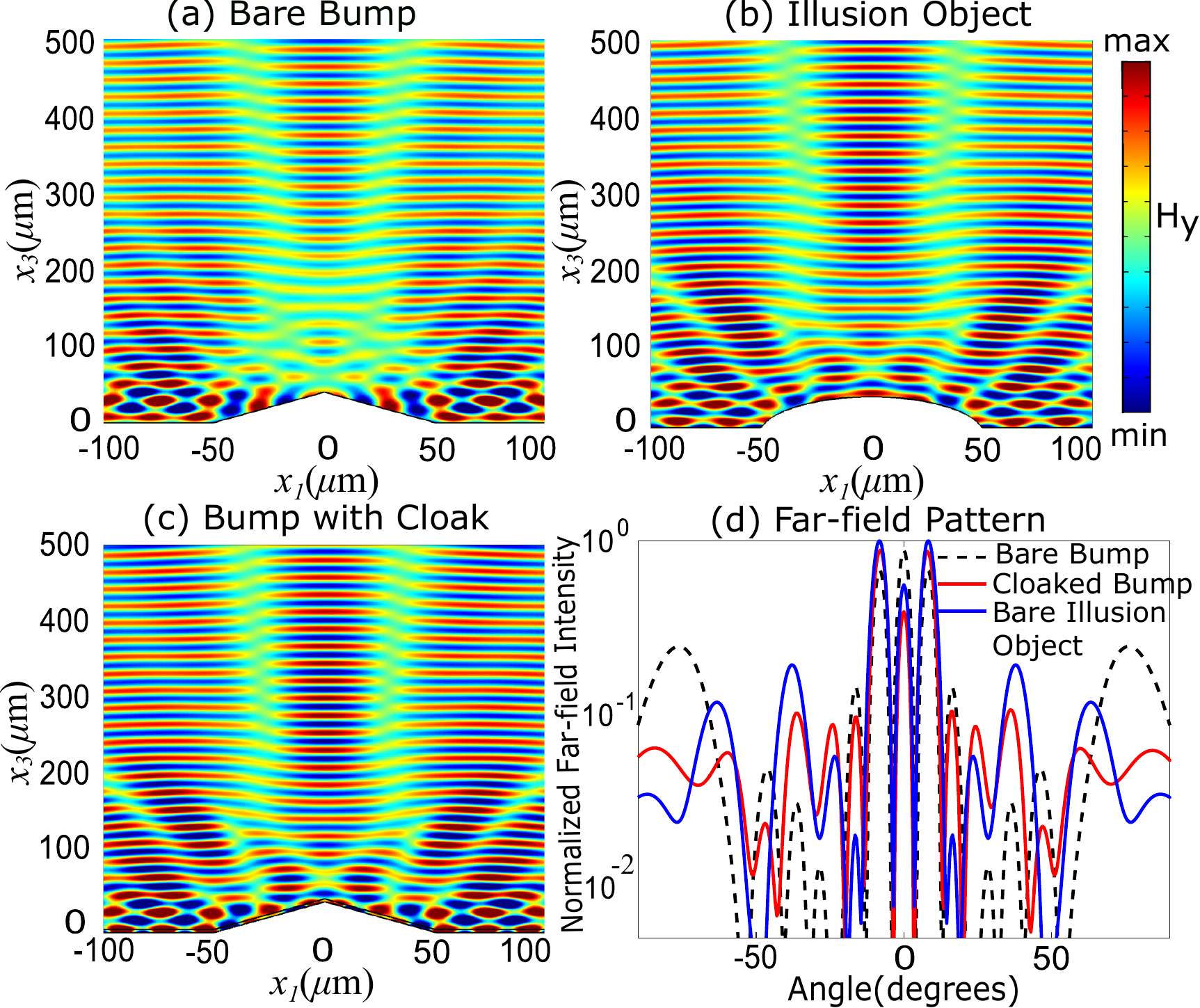}
	\caption{Simulation results for illusion optics. The cloaked object is a triangle shaped bump and the illusion object is a circular segment. (a),(b) and (c) show the scattered field plots for bare bump, bare illusion object and cloaked bump, while (d) shows corresponding far-field plots.}
	\label{fig:fig3}
\end{figure}

Fig.\,\ref{fig:fig3} shows the simulation results implementing the above mentioned $\phi$ for illusion optics. We have the same triangular bump as the object to be cloaked (i.e $\Gamma'$) and a circular segment with chord length 100$\mu$m and height 40$\mu$m as the desired illusion object i.e $\Gamma$. Fig.\,\ref{fig:fig3}a and \ref{fig:fig3}b show scattered field plots for the bare bump and illusion object respectively. When the triangular bump is cloaked by the designed metasurface, the scattering pattern becomes similar to that of the illusion object, which would make the triangular bump appear as a circular bump to an external observer. The field plot for the cloaked object is shown in Fig.\,\ref{fig:fig3}c. Angle resolved far-field intensity plots are shown in Fig.\,\ref{fig:fig3}d for comparison between these three cases. There is good agreement between the cloaked bump and illusion object in the far-field, especially within the angular window of $\pm 50^{\circ}$.

\section{Reflective Focusing} \label{sec5}

\textcolor{black}{In this section we consider focusing plane wave into a point $D(d_1,d_2,d_3)$ using metasurface parametrized by \eqref{Eq:point},\eqref{eq:normal}  (see Fig.\,\ref{fig:fig1}e). Assuming that $\bm{\alpha}$ is the constant unit incident vector,  we rewrite \eqref{eq:generalized law of reflection}  as
\[
\bm{\alpha}-\dfrac{D-P}{|D-P|}-\frac{\nabla \phi(P)}{k_0}=\lambda \,\bm{\nu} (P),
\]
where $\phi(P)$ is the phase discontinuity along the metasurface, and $(D-P)/{|D-P|}$ is the unit reflected vector.
Making the double cross product with $\bm{\nu}$ yields
\begin{align} \label{eq:vector_focus}
\frac{\nabla\phi(P)}{k_0}=\bm{\alpha}-\dfrac{D-P}{|D-P|}-\(\(\bm{\alpha}-\dfrac{D-P}{|D-P|}\)\cdot \bm{\nu}\)\,\bm{\nu}.
\end{align}
System of differential equations \eqref{eq:vector_focus} (see Appendix \ref{app:coord} for coordinate form) can be simplified by calculating derivatives of phase function, $\phi(P)$, with respect to $x_1$, $x_2$, using the chain rule (see \eqref{eq:chain_rule}),
\begin{align*}
\dfrac{\partial \phi }{\partial x_i}
&=k_0\(\alpha_i-\dfrac{d_i-x_i}{|D-P|}+\left(\alpha_3-\dfrac{d_3-g(X)}{|D-P|}\right) g_{x_i}(X)\)\\
&=k_0\(\dfrac{\partial}{\partial x_i} |D-P|+\dfrac{\partial}{\partial x_i} \(\alpha_i\,x_i+\alpha_3\,g(X)\)\)
\end{align*} }
with $i=1,2$. Therefore, we obtain the phase
\[
\phi(X,g(X))=k_0\(|D-P|+\bm{\alpha} \cdot \(X,g(X)\)\)+ C.
\]
For 2D geometry independent of $x_2$, $D=(x_d,z_d)$ and $\alpha=(-\sin\theta_i,-\cos\theta_i)$, the phase equation reduces to
\begin{equation}\label{eq:phase plane_to_point_2D}
\boxed{
\begin{aligned}
\phi(x_1)&=k_0\(\sqrt {{{(x_1 - {x_d})}^2} + {{(g(x_1) - {z_d})}^2}}-x_1\sin {\theta _i}\right.  \\
&- \left. g(x_1)\cos {\theta _i}\)+C.
\end{aligned}
}
\end{equation}

\textcolor{black}{In a similar way we can demonstrate (see Appendix \ref{app3}) that the phase discontinuity
\begin{equation}\label{eq:phase point_to_point_2D}
\boxed{
\begin{aligned}
\phi(x_1)&=k_0\(\sqrt {{{(x_1 - {x_s})}^2} + {{(g(x_1) - {z_s})}^2}}\right.  \\
&+ \left. \sqrt {{{(x_1 - {x_d})}^2} + {{(g(x_1) - {z_d})}^2}}\)+C
\end{aligned}
}
\end{equation}
should be imposed on the metasurface for focusing rays radiated by a point source located at $S = (x_s, z_s)$.  }

\begin{figure}[t]
\centering
\includegraphics[width=0.5\textwidth]{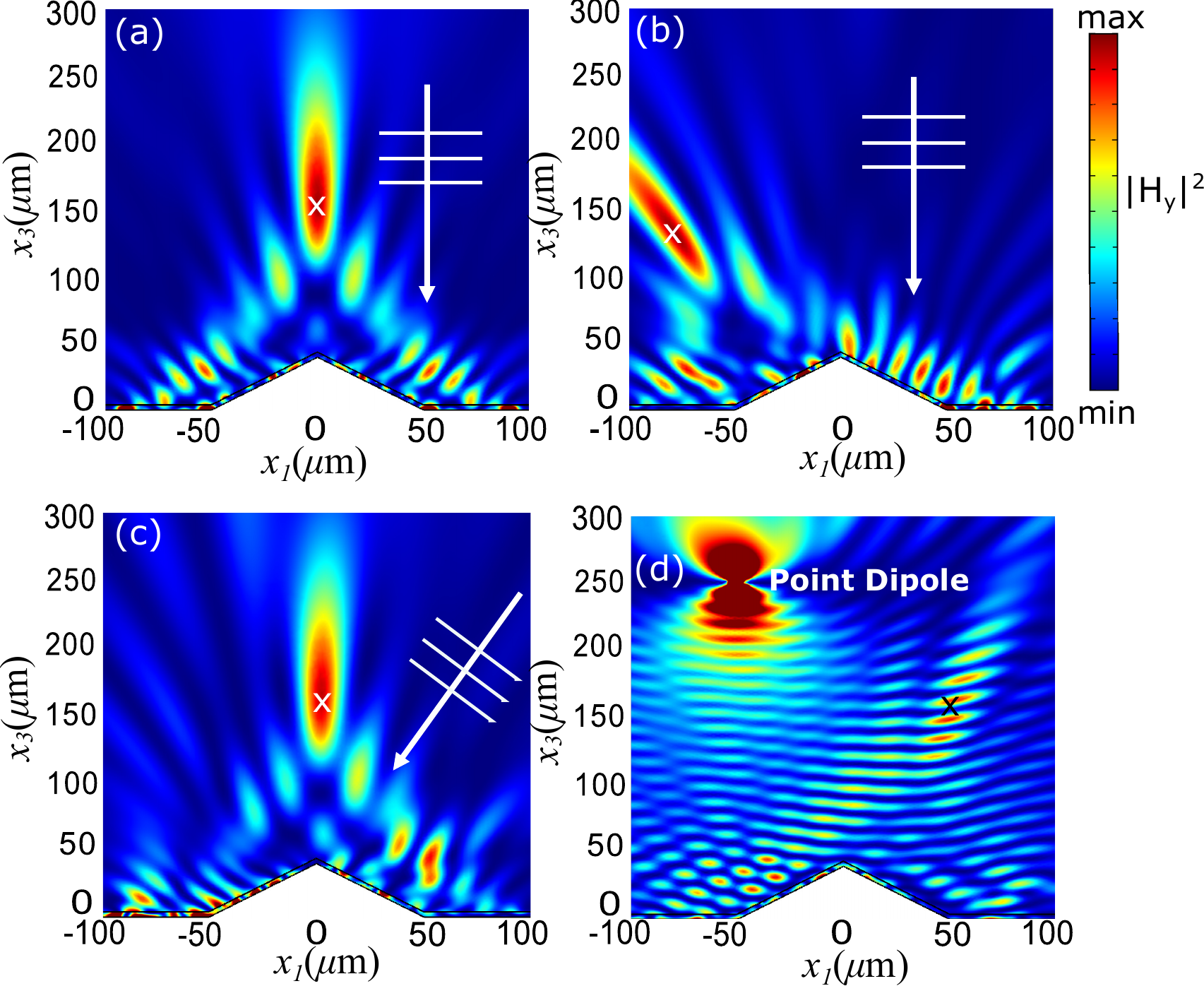}
\caption{Simulation results for reflective focusing off arbitrary surface. (a),(b) and (c) shows field intensity plots for focusing of an incident plane wave to a focal distance of 150$\mu$m from the ground plane. (a,b) show normal incidence while (c) shows result for $30^{\circ}$ incidence. In (a) and (c) the focal point is located 150$\mu$m away in normal direction while in (b) the focal point is at an angle of $30^{\circ}$. (d) shows focusing of a point dipole source.}
\label{fig:fig5}
\end{figure}

Simulation results for reflective focusing of incident parallel beams (plane wave) and point dipole source are shown in Fig.\,\ref{fig:fig5}. In scattered field intensity plots of Fig.\,\ref{fig:fig5}a, \ref{fig:fig5}b and \ref{fig:fig5}c, we have incident parallel beams focused to a point at a distance of 150$\mu$m from the base of the triangular bump (i.e ground plane). First we consider normal incidence. Fig.\,\ref{fig:fig5}a and Fig.\,\ref{fig:fig5}b shows simulation results for normal incidence. Lastly, we consider oblique incidence with $30^{\circ}$ angle in Fig.\,\ref{fig:fig5}c. Direction of incidence is shown by white arrows and position of the focusing point is indicated by `$\times$'. Flat gradient metasurfaces allow high numerical aperture (NA) diffraction limited focusing without spherical abberation\cite{aieta2012,yu2014}. The size of the focal spot in Fig.\,\ref{fig:fig5}a, b and c is comparable to the free space wavelength of 22$\mu$m.

In Fig.\,\ref{fig:fig5}d, an  example of focusing of a point source is shown. The source is at (-50,250)$\mu$m and focusing point is at (50,150)$\mu$m. The point source is modeled by a electric point dipole in COMSOL with its dipole moment oriented along $x_1$-direction. As there is no straight forward way to use a point source for scattered field calculation in COMSOL, we simulate for the total field instead with a point dipole acting as a point source. The plotted quantity in Fig.\,\ref{fig:fig5}d, is the total field intensity i.e both incident and reflected fields are present. We can see higher intensity of field around the designed focus point indicating the focusing effect.

\section{Discussion and Conclusion}
Concluding, we demonstrated versatility of graphene-based metasurface that is capable of active switching between regimes of operation such as anomalous beam steering, focusing, cloaking and illusion optics simply by changing electric bias applied to graphene constituents of the metasurface without changing the metasurface geometry. \textcolor{black}{These various functionalities are usually available in a disparate manner in the existing literature and we showed in this work that they can be described within a general framework for arbitrary surface morphology. The proposed approach, in particular to the context of graphene metasurface, makes perfect sense since graphene can be electrically tunable to achieve arbitrary phase function and conform to any surface morphology.} As an example, we considered triangular bump covered by the graphene metasurface, made from graphene ribbons on the dielectric resonator, and demonstrated that, by applying an electric bias, the wavefront of the wave reflected by the bump can be tuned to match that of the bare plane (cloaking) or hemi-sphere (illusion optics). Moreover, the possibility of anomalous steering and focusing of the wave reflected by graphene metasurface covered bump was shown. The slight distortion of the metasurface far-field radiation pattern from that of the bare plane or hemi-sphere can be attributed to the specular reflection from the parts of metasurface not-covered by graphene ribbons, as well as to the fact that reflectivity of graphene ribbon depends on the applied electric bias. \textcolor{black}{Finite size effects also show up in the field profile due to finiteness of simulation domain, discretization of metasurface and contribution from the apex of the triangle\cite{wei2017ultrathin,orazbayev2015}}.  We expect that by optimizing the metasurface geometry these distortions can be reduced.

The device configuration considered here can be fabricated with conventional film deposition and nanopatterning technologies. The transfer of graphene\cite{lee2010wafer,martins2013direct} onto bump structure and its patterning by electron beam lithography would be straightforward as demonstrated elsewhere\cite{wang2011super,hofmann2014scalable}. Nevertheless, there are a few issues that need to be addressed in terms of practical implementation. First of all, we should select a proper material for optical spacer, which is transparent over the concerned frequency range and compatible with conventional thin film deposition technologies. In addition, it is important to have small roughness on the film surface for the graphene transfer that follows. For mid-infrared applications, silicon oxide (SiO$_2$)\cite{rodrigo2015mid} and hexagonal boron nitride (hBN)\cite{dai2015graphene} have been popularly used as substrates for graphene, although plasmon losses due to strong plasmon-phonon coupling should be taken into consideration to determine the operation wavelength. Diamond-like carbon\cite{yan2013damping} and calcium fluoride (CaF$_2$)\cite{hu2016far} can be good candidates as they do not have polar phonons in this frequency range. The issue of graphene and substrate losses are discussed in Appendix \ref{app2}. The insulating property and dielectric strength of the material used for the optical spacer becomes one of the important design parameters, from which the tunable range of graphene conductivity is largely determined. Another important aspect is addressing individual ribbons for separate doping. Since with other types of doping using chemical vapor and ion gel it is difficult to address individual ribbons, electrical doping by separate electrodes is necessary for the device. \textcolor{black}{A recent work\cite{barik2017graphene} demonstrates that embedded local gating structures with graphene is experimentally feasible.}
\begin{acknowledgments}
This work is supported by a DARPA grant award FA8650-16-2-7640 and partially supported by NSF grant DMS-1600578.
\end{acknowledgments}

\appendix
\section{Generalized Snell's Law in Vector Form}\label{app1}
Let rays of light be incident from point $S=(s_1,s_2,s_3)$, at a point $P(x_1,x_2,x_3=a)$ on a plane parallel to the $x_1-x_2$ plane, located at $x_3=a$. Incident rays are then reflected to point $D=(d_1,d_2,d_3)$. The normal to $P$ is $\bm{\nu}=\hat{k}\equiv(0,0,1)$.
Therefore the incident unit vector from $S$ into a point $P$ on $\Gamma$ is
$\bm \alpha=\dfrac{\overrightarrow{SP}}{|\overrightarrow{SP}|}$ and the reflected unit vector from $P$ into $D$ is
$\bm \beta=\dfrac{\overrightarrow{PD}}{|\overrightarrow{PD}|}$.
Since the ray is propagating in vacuum, from Fermat's principle, the least optical paths for the incident and reflected rays are given by $\left|\overrightarrow{SP}\right|$ and $\left|\overrightarrow{DP}\right|$ and the corresponding accumulated phases are given by
$k_0\,\left|\overrightarrow{SP}\right|$ and $k_0\,\left|\overrightarrow{DP}\right|$ respectively; where $k_0$ is the free space wave number and $|\cdot |$ denotes the Euclidean distance.
We introduce a phase discontinuity $\phi$ along $\Gamma$. According to principle of stationary phase\citep{aieta2012out,genSnell}, we then seek to minimize the total phase $k_0\,\left|\overrightarrow{SP}\right|+k_0\,\left|\overrightarrow{DP}\right|-\phi(P)$ for $P\equiv(x_1,x_2,a)$ in $\Gamma$.
Therefore at the extreme point on $\Gamma$, by differentiating the total phase with respect to $x_1$ and $x_2$, we must have
\begin{align*}
k_0\,\dfrac{x_1-s_1}{\left|\overrightarrow{SP}\right|}+k_0\,\dfrac{x_1-d_1}{\left|\overrightarrow{DP}\right|}&=\dfrac{\partial \phi}{\partial x_1}\\
k_0\,\dfrac{x_2-s_2}{\left|\overrightarrow{SP}\right|}+k_0\,\dfrac{x_2-d_2}{\left|\overrightarrow{DP}\right|}&=\dfrac{\partial \phi}{\partial x_2}
\end{align*}
which from the definitions above of $\bm \alpha$ and $\bm \beta$ can be rewritten as
\begin{align*}
\left( k_0\,\bm \alpha -k_0\,\bm \beta\right)\cdot \hat{ i}=\dfrac{\partial \phi}{\partial x_1}\qquad
\left( k_0\,\bm \alpha -k_0\,\bm \beta\right)\cdot \hat{ j}=\dfrac{\partial \phi}{\partial x_2},
\end{align*}
for $x_1,x_2$ and $x_3=a$.
Since the normal $\bm{\nu}=\hat{k}$, we therefore obtain
the following expression of the generalized reflection law:
\[
k_0\,\bm \alpha -k_0\,\bm \beta=\dfrac{\partial \phi}{\partial x_1}\,\hat{i}
+
\dfrac{\partial \phi}{\partial x_2}\,\hat{j} +\xi\, \bm{\nu}.
\]
Notice that when there is no phase discontinuity,  i.e. $\phi=0$, we recover the standard reflection law in vector form.
If $\phi$ is defined in a small neighborhood of the plane $\Gamma$, i.e. $\phi(x_1,x_2,x_3)$ is defined for all $x_1,x_2$ and for $x_3-a$ very small, then we can write the formula
\textcolor{black}{
\begin{equation}\label{eq:generalizedSnell}
\bm{\alpha}-\bm{\beta}=\frac{\nabla \phi}{{{k_0}}}+\lambda\,\bm{\nu}
\end{equation}}
where $\lambda$ is a scalar.
\begin{figure}[t]
\centering
\includegraphics[width=0.3\textwidth]{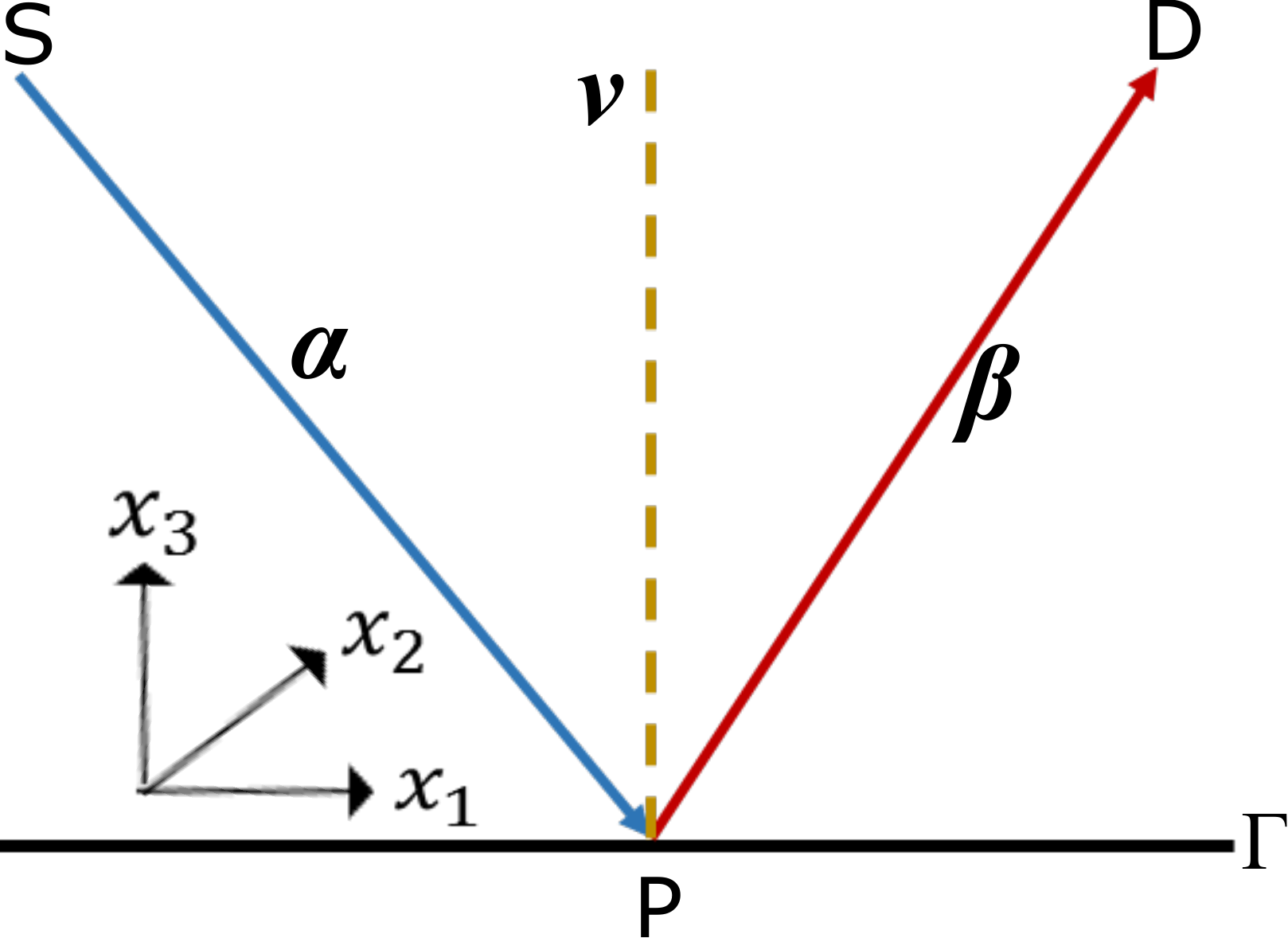}
\caption{Ray diagram illustrating generalized Snell's law, see also Eq.\,\eqref{eq:generalizedSnell} }
\label{fig:fig6}
\end{figure}
Eq.\,\eqref{eq:generalizedSnell} is the vector form of generalized Snell's law for reflection.

\section{Effect of Loss on Phase}\label{app2}
\begin{figure}[t]
\centering
\includegraphics[width=0.35\textwidth]{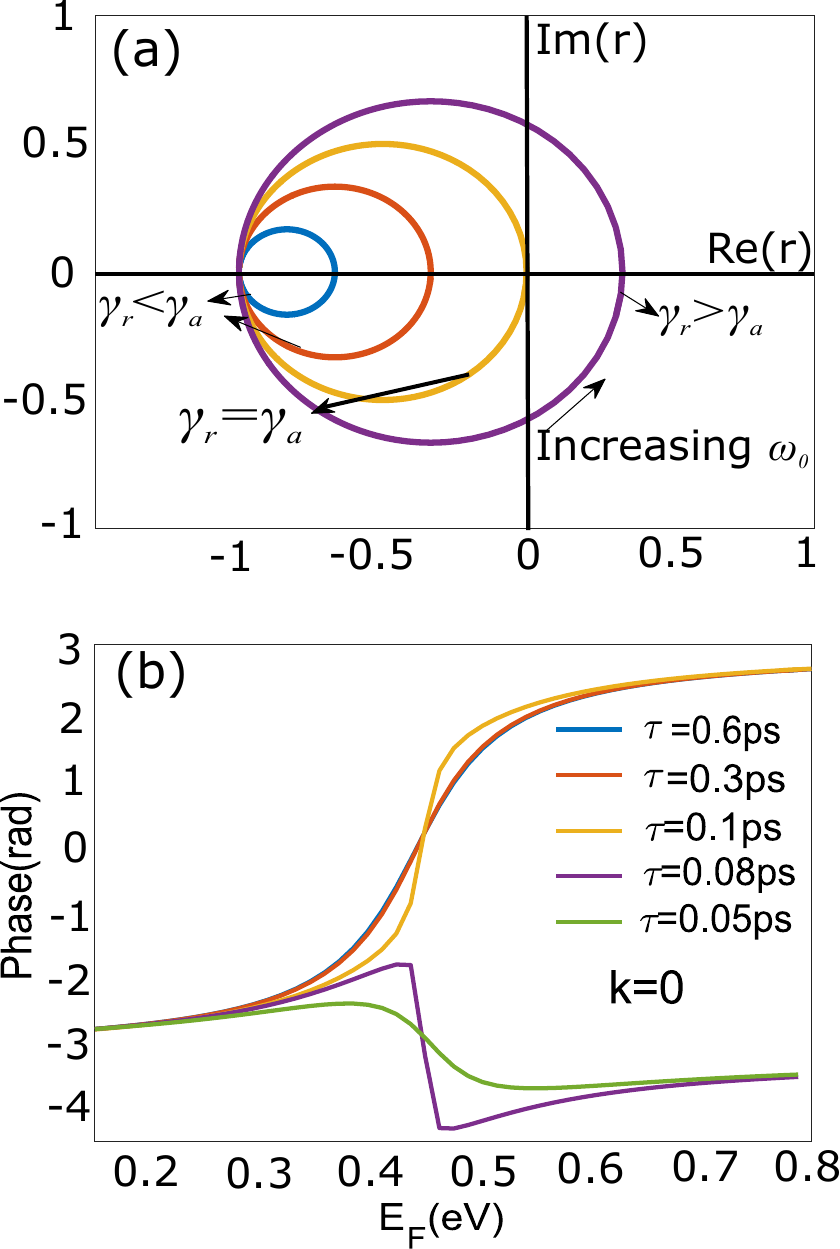}
\caption{Effect of intrinsic losses on achievable phase range. (a) shows complex plane plot of reflection coefficient for an analytical model of the structure as described in \citep{hauswaves} using single port resonator model. When intrinsic losses exceed external or radiative losses, phase of $r$ cannot cover all four quadrants (blue and red curves) of the complex plane and $2\pi$ phase shift is lost. (b) shows simulation results of $E_F$ vs. Phase for the device for different relaxation times, $\tau$. As $\tau$ goes below 0.1ps, we see a drastic change in phase profile.}
\label{fig:loss}
\end{figure}
The attainable range of reflection phase is dependent on absorptive losses in the device. The reason for losing phase shift of $2\pi$ with increased losses can be explained with arguments based on coupled mode theory (CMT).
The device structure of graphene-substrate-metal creates an asymmetric Fabry-Perot resonator with a perfectly reflective mirror (metal) and a partially reflective mirror (graphene-dielectric layer interface). This can be effectively described as a one port single resonator, working at a resonant frequency of $\omega_0$\citep{effectloss}. According to CMT, when the resonator is excited by an external excitation of frequency $\omega$, the reflection coefficient is given by\citep{hauswaves}
\[r = \frac{{{\gamma _r} - {\gamma _a} - i\left( {\omega  - {\omega _0}} \right)}}{{{\gamma _r} + {\gamma _a} + i\left( {\omega  - {\omega _0}} \right)}}.\]

Where $\gamma_r=1/\tau_r$ is the rate of external or radiative losses and $\gamma_a=1/\tau_a$ is the rate of internal of absorptive losses. Fig.\,\ref{fig:loss}a shows the plot of $r$ in complex plane for different $\omega_0$ with fixed $\omega$. As can be seen on the plot, when absorptive losses are smaller than radiative losses ($\gamma_r>\gamma_a$), $r$ covers all four quadrants in the complex plane and reflection phase covers the whole $-\pi$ to $\pi$ range. This situation is called underdamped. But when absorptive loss surpasses radiative loss i.e $\gamma_r<\gamma_a$, the phase of $r$  can no longer go from $-\pi$ to $\pi$ and the system is called overdamped.

In our device, by changing $E_F$, the plasmon resonance frequency is varied, as $\omega_0\propto\sqrt{(E_F)}$. The radiative losses ($\gamma_r$) are constant as they are dependent on the dimensions of the device. The absorptive losses ($\gamma_a$) are proportional to inverse of relaxation time, $1/\tau$, and imaginary part of refractive index of the dielectric cavity, $k$. Hence when $\tau$ is decreased or $k$ is increased, the system moves from underdamped to overdamped and the $2\pi$ phase shift range is lost.
In Fig.\,\ref{fig:loss}b, reflection phase is plotted as a function of $E_F$ for different relaxation times $\tau$. The phase shift range becomes much smaller than $2\pi$ when $\tau$ is decreased below $~0.1$ps. Similar behavior can be seen when $k$ is increased above $0.15$. Both $\tau$ and $k$ are parameters related to total absorptive losses in the device.
A similar phase behavior was observed in \citep{effectloss} for metal-insulator-metal (MIM) based metasurfaces.
\section{\textcolor{black}{Coordinate representation of the vector equations for phase function }} \label{app:coord}

\textcolor{black}{
Coordinate form of the vector equation \eqref{eq:vector_cloaking}
\begin{align*}
\dfrac{\partial \phi(P)}{\partial u_1}&=k_0\(\alpha_1-\beta_1+\delta\,  g_{x_1}(X)\)\\
\dfrac{\partial \phi(P)}{\partial u_2}&=k_0\(\alpha_2-\beta_2+\delta\, g_{x_2}(X)\)\\
\dfrac{\partial \phi(P)}{\partial u_3}&=k_0\(\alpha_3-\beta_3-\delta\).
\end{align*}}

\textcolor{black}{
Coordinate form of the vector equation \eqref{eq:gradient}
\begin{align*}
\dfrac{\partial \phi(P)}{\partial u_1}
&=
 2 k_0\(\bm{\alpha}\cdot \bm{\eta}(P')\) \,\left( \dfrac{-f_{x_1}(X)}{\sqrt{1+|\nabla f(X)|^2}}\right.\\
&+\left. \(\bm{\eta}(P')\cdot \bm{\nu}(P)\)\, \dfrac{g_{x_1}(X)}{\sqrt{1+|\nabla g(X)|^2}} \right)\\
\dfrac{\partial \phi(P)}{\partial u_2}
&=
 2 k_0\(\bm{\alpha}\cdot \bm{\eta}(P')\) \,\left(\dfrac{-f_{x_2}(X)}{\sqrt{1+|\nabla f(X)|^2}}\right.\\
&+ \left.\(\bm{\eta}(P')\cdot \bm{\nu}(P)\)\, \dfrac{g_{x_2}(X)}{\sqrt{1+|\nabla g(X)|^2}} \right)\\
\dfrac{\partial \phi(P)}{\partial u_3}
&=
 2k_0\(\bm{\alpha}\cdot \bm{\eta}(P')\) \,\left(\dfrac{1}{\sqrt{1+|\nabla f(X)|^2}}\right.\\
&- \left.\(\bm{\eta}(P')\cdot \bm{\nu}(P)\)\, \dfrac{1}{\sqrt{1+|\nabla g(X)|^2}} \right).
\end{align*}}

\textcolor{black}{Coordinate form of the vector equation \eqref{eq:vector_focus}
\begin{align*}
\dfrac{\partial \phi(P)}{\partial u_1}&=k_0\(\alpha_1-\dfrac{d_1-x_1}{|D-P|}\right.\\
&+\left.\(\(\bm{\alpha}-\dfrac{D-P}{|D-P|}\)\cdot \bm{\nu}\)\,\dfrac{g_{x_1}(X)}{\sqrt{1+|\nabla g(X)|^2}}\)\\
\dfrac{\partial \phi(P)}{\partial u_2}&=k_0\(\alpha_2-\dfrac{d_2-x_2}{|D-P|}\right.\\
&+\left.\(\(\bm{\alpha}-\dfrac{D-P}{|D-P|}\)\cdot \bm{\nu}\)\,\dfrac{g_{x_2}(X)}{\sqrt{1+|\nabla g(X)|^2}}\)\\
\dfrac{\partial \phi(P)}{\partial u_3}&=k_0\(\alpha_3-\dfrac{d_3-g(X)}{|D-P|}\right.\\
&-\left.\(\(\bm{\alpha}-\dfrac{D-P}{|D-P|}\)\cdot \bm{\nu}\)\,\dfrac{1}{\sqrt{1+|\nabla g(X)|^2}}\).
\end{align*}}

\section{\textcolor{black}{Integrating of a system of differential equations \eqref{eq:partial xi psi general}}} \label{app:integrate}
\textcolor{black}{In this section we integrate a system of differential equations, \eqref{eq:partial xi psi general}, obatined for the illusion optics case,
\begin{align*}
\dfrac{\partial \phi}{\partial x_1}
&=
 2k_0\(\bm{\alpha}\cdot \bm{\eta}(P')\)\,\dfrac{g_{x_1}(X)-f_{x_1}(X)}{\sqrt{1+|\nabla f(X)|^2}}:=A_1(X) \\
\dfrac{\partial \phi}{\partial x_2}
&= 2k_0\(\bm{\alpha}\cdot \bm{\eta}(P')\)\,\dfrac{g_{x_2}(X)-f_{x_2}(X)}{\sqrt{1+|\nabla f(X)|^2}} :=A_2(X)
\end{align*}}

If $\phi$ and $g$ are $C^2$, then the mixed partials $\dfrac{\partial^2 }{\partial x_1\partial x_2}\(\phi(X,g(X))\)$
and $\dfrac{\partial^2 }{\partial x_2\partial x_1}\(\phi(X,g(X))\)$ must be equal.
Therefore to have a solution $\phi$ the following compatibility condition between $\bm{\alpha}$, $f$ and $g$ must hold:
\begin{equation}\label{eq:compatibility condition}
\dfrac{\partial }{\partial x_2}A_1(X)
=
\dfrac{\partial }{\partial x_1}A_2(X).
\end{equation}
In fact, if \eqref{eq:compatibility condition} holds we will obtain the phase $\phi$ by integration as follows.
To simplify the notation, set $h(X)=\phi\(X,g(X)\)$, so we need to solve the system
\[
\dfrac{\partial h}{\partial x_1}=A_1,\qquad \dfrac{\partial h}{\partial x_2}=A_2.
\]
Integrating the first equation with respect to $x_1$ yields
\[
h(x_1,x_2)=\int_a^{x_1}A_1(s,x_2)\,ds+W(x_2).
\]
Differentiating the last equation with respect to $x_2$ gives
\begin{align*}
\dfrac{\partial h}{\partial x_2}(x_1,x_2)&=\int_a^{x_1}\dfrac{\partial A_1}{\partial x_2}(s,x_2)\,ds+W'(x_2)\\
&=\int_a^{x_1}\dfrac{\partial A_2}{\partial x_1}(s,x_2)\,ds+W'(x_2)\qquad \text{from \eqref{eq:compatibility condition}}\\
&=
A_2(x_1,x_2)-A_2(a,x_2)+W'(x_2).
\end{align*}
So $W'(x_2)=A_2(a,x_2)$, and by integration $W(x_2)=\int_b^{x_2}A_2(a,t)\,dt +C$.
Therefore, we obtain
\begin{equation*}
\phi\(X,g(X)\)=\int_a^{x_1}A_1(s,x_2)\,ds+\int_b^{x_2}A_2(a,t)\,dt +C.
\end{equation*}

\section{\textcolor{black}{Focusing from point source to point}} \label{app3}
\textcolor{black}{Here we} devise a metasurface for reflective focusing due to a point source. Let $S(s_1,s_2,s_3)$ and $D(d_1,d_2,d_3)$ be two points above the surface \textcolor{black}{parameterized by \eqref{Eq:point}, \eqref{eq:normal}}. We seek a phase discontinuity so that all rays incident from $S$ are reflected into $D$.
Then the incident unit direction equals $\dfrac{P-S}{|P-S|}$ and the reflected unit direction equals $\dfrac{D-P}{|D-P|}$.
From \eqref{eq:generalized law of reflection} we then seek $\phi$ so that
\[
\dfrac{P-S}{|P-S|}-\dfrac{D-P}{|D-P|}-\frac{\nabla \phi(P)}{k_0}=\lambda \,\bm{\nu} (P).
\]
Following similar steps as discussed \textcolor{black}{in Section \ref{sec5}},
\textcolor{black}{
\begin{align*}
\frac{\nabla \phi(P)}{k_0} &=\dfrac{P-S}{|P-S|}-\dfrac{D-P}{|D-P|}\\
& -\(\(\dfrac{P-S}{|P-S|}-\dfrac{D-P}{|D-P|}\)\cdot \bm{\nu}\)\,\bm{\nu}.
\end{align*}
Writing in coordinates yields
\begin{align*}
\phi_{u_1}&=k_0\(\dfrac{x_1-s_1}{|P-S|}-\dfrac{d_1-x_1}{|D-P|}\right.\\
&+\left.\(\(\dfrac{P-S}{|P-S|}-\dfrac{D-P}{|D-P|}\)\cdot \bm{\nu}\)\,\dfrac{g_{x_1}(X)}{\sqrt{1+|\nabla g(X)|^2}}\)\\
\phi_{u_2}&=k_0\(\dfrac{x_2-s_2}{|P-S|}-\dfrac{d_2-x_2}{|D-P|}\right.\\
&+\left.\(\(\dfrac{P-S}{|P-S|}-\dfrac{D-P}{|D-P|}\)\cdot \bm{\nu}\)\,\dfrac{g_{x_2}(X)}{\sqrt{1+|\nabla g(X)|^2}}\)\\
\phi_{u_3}&=k_0\(\dfrac{g(X)-s_3}{|P-S|}-\dfrac{d_3-g(X)}{|D-P|}\right.\\
&-\left.\(\(\dfrac{P-S}{|P-S|}-\dfrac{D-P}{|D-P|}\)\cdot \bm{\nu}\)\,\dfrac{1}{\sqrt{1+|\nabla g(X)|^2}}\).
\end{align*}}
Hence by the chain rule
\textcolor{black}{
\begin{align*}
\dfrac{\partial \phi}{\partial x_i}
&=\dfrac{\partial \phi(u_1,u_2,u_3)}{\partial u_i}+\dfrac{\partial \phi(u_1, u_2, u_3)}{\partial u_3}\frac{\partial u_3}{\partial x_i}\\
&=k_0\(\dfrac{x_i-s_i}{|P-S|}-\dfrac{d_i-x_i}{|D-P|}\right.\\
&+\left.\dfrac{g(X)-s_3}{|X-A|}\,g_{x_i}(X)-\dfrac{d_3-g(X)}{|D-P|}\,g_{x_i}(X)\)\\
&=k_0\(\dfrac{\partial}{\partial x_i} |P-S|
+\dfrac{\partial}{\partial x_i} |D-P|\)
\end{align*}}
with $i=1,2$. Therefore we obtain the phase as
\[
\phi=k_0\(|P-S|+|D-P|\)+C.
\]
For 2D geometry independent of $x_2$, $S=(x_s,z_s)$ and $D=(x_d,z_d)$, the phase equation reduces to
\begin{multline}
\phi(x_1)=k_0\(\sqrt {{{(x_1 - {x_s})}^2} + {{(g(x_1) - {z_s})}^2}}\right.  \\
+ \left. \sqrt {{{(x_1 - {x_d})}^2} + {{(g(x_1) - {z_d})}^2}}\)+C.
\end{multline}

\providecommand{\noopsort}[1]{}\providecommand{\singleletter}[1]{#1}%
%

\end{document}